\documentstyle[aps,preprint]{revtex}

\begin{document}

\preprint{\vbox{\hbox{ }}}
\title{Equivalent Relations between Quantum Dynamics as Derived from
       a Gauge Transformation}
\author{Chuan-Tsung Chan\thanks{email address: ctchan@phys.ntu.edu.tw},
        Chun-Khiang Chua\thanks{email address: ckchua@phys.ntu.edu.tw}}
\address{Department of Physics, National Taiwan University, Taipei,
         Taiwan, 10764}
\date{\today}
\maketitle
\begin{abstract}
  Equivalent relations between quantum mechanical systems in the
Robertson-Walker (RW) background metric and quantum dynamics with
an induced quadratic background potential are derived in this work.
  Two elementary applications, which include an algebraic derivation
of the evolution operator for a simple harmonic oscillator without
using any special function or the path integral technique, and a
moving soliton solution of a free particle in an oscillating universe, 
are presented to illustrate the use of these equivalent relations.
\end{abstract}

\vspace{2cm}

\pacs{PACS numbers:}

\newpage

\section{Motivation and Introduction}

  Recent advance of duality in string theory has not only greatly
improved our understanding of string theory as an unified framework of
quantum gravity, but also drastically changed our view of quantum field
theory.
  For instance, the Maldacena conjecture gives an unforeseen correspondence 
between super-gravity and super-Yang-Mills theory which allows us to study 
the strongly coupled dynamics of gauge theories via black hole physics 
\cite{mad}.
  Another example regarding strong-weak duality which exchanges fundamental 
with solitonic degrees of freedom in apparently different models and 
establishes their equivalence \cite{Duff}.

  On the other hand, it has also become clear that, as a technical tool,
the use of duality facilitates investigations of nonperturbative or
strong coupling aspects in quantum dynamics, which are certainly beyond
the domain of perturbative calculations.
  While there exists vast literature providing substantial evidences for
supporting various duality relations as exact quantum symmetries, it remains 
to be a great theoretical challenge to find mathematical derivations of these 
conjectured connections.

  In this work, we shall focus on a special duality transformation which
generates equivalent relations between quantum systems with different 
backgrounds.
  Due to relatively simple structures of these systems, it is possible
to derive these equivalent relations from a series of change of variables.
  The precise correspondence between two classes of quantum systems can be 
formulated as transformations between wave functions and evolution operators, 
and we can explore physical consequences based on such an equivalent relation.

  This paper is organized as follows.
  We first derive the equivalent relations between quantum systems in the
Robertson-Walker metric (denoted as Class A) and quantum systems with an 
induced quadratic background potential (denoted as Class B) in Sec.II.
  The use of these equivalent relations is illustrated by two elementary
applications in Sec.III and Sec.IV.
  In particular, we find an algebraic derivation of evolution operator for a 
non-relativistic simple harmonic oscillator and a moving solitonic wave of a 
free particle in the space with oscillating metric.
  After making comparison with other approaches and commenting on special
features of this equivalent relation in Sec.V, we summarize and conclude 
this paper in Sec.VI.

\section{Equivalent Relations between Quantum Dynamics with Different 
         Backgrounds}

  The Schr\"{o}dinger equation in the Robertson-Walker (RW) metric \cite{cos},
\begin{equation}
  d s^2 = g_{\mu \nu} d x^\mu d x^\nu
= d t^2 - a^2 (t) \left[ \frac{d r^2}{1 - k r^2}
  + r^2 d \theta^2 + r^2 \sin^2 \theta d \phi^2  \right],
\label{eq:rw}
\end{equation}
is given by \cite{pqm}
\begin{eqnarray}
 i \frac{\partial}{\partial t} \psi(\vec x, t) &=&
 \left[ - \frac{1}{2 m \sqrt{g}} \sum_{i,j}
 \frac{\partial}{\partial x^i} \left( \sqrt{g} g^{ij}
 \frac{\partial}{\partial x^j} \right)
 + {\cal V} ( \vec x, t; g_{\mu\nu} ) \right] \psi(\vec x, t) \nonumber \\
 &=& \left[ - \frac{ {\nabla}_x^2 }{2 a^2 (t) m}
 + V ( a(t) \vec x, t ) \right] \psi(\vec x, t),
\label{eq:sch1}
\end{eqnarray}
where $\nabla_x^2$ is the Laplacian in the space of constant curvature $k$,
and $a(t)$ is the scale factor describing the change of metric under the 
dynamics of general relativity.
  In the present study, we shall confine ourselves to the case of flat
space, i.e., $k=0$.

  To solve this time-dependent Schr\"{o}dinger equation, we observe that 
the scale factor $a(t)$ appearing in the Hamiltonian through a rescaling of 
the spatial coordinates $\vec x$, and it is tempting to redefine new 
coordinates and the wave function, 
\begin{equation}
  \vec r \equiv a(t) \vec x, \hspace{1cm}
  f (\vec r, t) \equiv \psi \left( \frac{\vec r}{a(t)}, t \right),
\label{eq:change1}
\end{equation}
to absorb such a time-dependent scale factor.
  Nevertheless, one needs to be careful in transforming the time derivative 
on the left hand side of the Schr\"{o}dinger equation, Eq.(\ref{eq:sch1}), 
for now we have to fix new spatial variable $\vec r$ while taking the partial 
derivative with respect to time variable, 
\begin{equation}
  i \frac{\partial}{\partial t} \psi ( \vec x, t)
  \left|_{\vec x} \right.
= i \left[ \frac{\dot a (t)}{a (t)} \right] \ \vec r \cdot
    \left( \frac{\partial f}{\partial \vec r} \left|_t \right. \right)
+ i \frac{\partial f}{\partial t} \left|_{\vec r} \right. .
\end{equation}
  In terms of these new variables, the Schr\"{o}dinger equation,  
Eq.(\ref{eq:sch1}), becomes 
\begin{equation}
        i  \frac{\partial}{\partial t} f (\vec r, t) =
  \left[ - \frac{ {\nabla_r}^2 }{2m} - i h(t) \left( \vec r \cdot
  \nabla_r \right) + V( \vec r, t) \right] f (\vec r, t),
\label{eq:sch2}
\end{equation}
where we have introduced the Hubble parameter \cite{cos},
\begin{equation}
h(t) \equiv \frac{\dot a (t)}{a (t)}.
\end{equation}

  The non-Hermitian term, $- i h(t) \left( \vec r \cdot \nabla_r \right)$,
in the Hamiltonian, Eq.(\ref{eq:sch2}), originates from the time-dependent 
normalization of the new wave function, 
\begin{equation}
   \int d^3 x \ | \psi(\vec x, t) |^2 = 1
   \hspace{0.75cm} \Rightarrow \hspace{0.75cm}
   \int d^3 r \ |   f (\vec r, t) |^2 = | a (t) |^3.
\end{equation}
  In view of this, one can recover a Hermitian Hamiltonian by redefining a
properly normalized wave function,
\begin{equation}
   g (\vec r, t) \equiv a^{-3/2} (t) \ f (\vec r, t)
   \hspace{0.75cm} \Rightarrow \hspace{0.75cm}
  \int d^3 r \ | g (\vec r, t) |^2 = 1,
\label{eq:change2}
\end{equation}
and the Schr\"{o}dinger equation becomes
\begin{equation}
         i  \frac{\partial}{\partial t} g (\vec r, t) =
  \left[ - \frac{\nabla_r^2}{2 m}
         - \frac{i}{2} h(t) \left\{ \vec r, \nabla_r \right\}
         + V( \vec r, t )
 \right] g (\vec r, t).
\label{eq:sch3}
\end{equation}
  In this representation, the induced potential, which is in proportional to 
$ \left\{ \vec r, \nabla_r \right\} \equiv
          \vec r \cdot \nabla_r + \nabla_r \cdot \vec r $, 
suggests a form of gauge coupling.
  That is, we can rewrite the Hamiltonian as
\begin{equation}
 H(t) = \frac{1}{2 m}
        \left[ - i {\nabla}_r + m h(t) \vec r \ \right]^2
      - \left( \frac{m}{2} \right) h^2(t) {\vec r}^{\ 2} + V( \vec r, t ),
\end{equation}
and make further simplification by employing a gauge transformation,
\begin{equation}
    \phi (\vec r, t) \equiv
    \exp \left[ i \frac{m}{2} h(t) {\vec r}^{\ 2} \right]
    g (\vec r, t).
\label{eq:gauge1}
\end{equation}
  With all these changes of variables,
Eqs.(\ref{eq:change1}), (\ref{eq:change2}), (\ref{eq:gauge1}), we
generate a ``dual" quantum system with an induced quadratic potential,
\begin{equation}
         i \frac{\partial}{\partial t} \phi (\vec r, t) =
  \left[ - \frac{ {\nabla_r}^2 }{2m}
         - \frac{m}{2} \left( \frac{\ddot a}{a} \right) {\vec r}^{\ 2}
         + V( \vec r, t) \right]
         \phi (\vec r, t).
\label{eq:sch4}
\end{equation}

  Comparing the original Schr\"{o}dinger equation, Eq.(\ref{eq:sch1}),
to its dual representation, Eq.(\ref{eq:sch4}), we find that the 
time-dependent kinetic term is replaced by the standard form and the
effect of the Robertson-Walker background metric manifests itself as a
quadratic potential.
  Note that we need nontrivial time-dependence of the scale factor,
$\ddot a (t) \not = 0$, to generate an induced potential.
  Furthermore, while it seems unclear how to search for experimental
consequences of the systems, Eq.(\ref{eq:sch1}), other than its cosmology 
context; the dual systems, Eq.(\ref{eq:sch4}), could find their experimental 
realization easily in a laboratory environment.

  Given a scale factor $a(t)$, our derivation establishes an equivalent
relation between quantum systems in a time-dependent background metric, 
Eq.(\ref{eq:sch1}), and quantum systems with an induced quadratic potential, 
Eq.(\ref{eq:sch4}).
  To be precise, we have the following correspondence between wave functions,
\begin{eqnarray}
 \psi (\vec x, t) &=& a^{3/2} (t) \exp \left[ - i \ \frac{m}{2} \
 \dot a(t) \ a(t) \ {\vec x}^{\ 2} \right] \phi (a(t) \vec x, t),
\label{eq:dual1} \\
 \phi (\vec r, t) &=& a^{- 3/2} (t)
 \exp \left[ i \ \frac{m}{2} h(t) \ {\vec r}^{\ 2} \right]
 \psi \left( \frac{\vec r}{a(t)}, t \right).
\label{eq:dual2}
\end{eqnarray}
  Notice that, apart from the scale factor, which ensures proper normalization,
Eq.(\ref{eq:change2}), the correspondence between wave functions in two classes
of systems is given by a time-dependent gauge transformation.
  Given the transformation rules between wave functions, we can deduce that 
the evolution operators for each class of systems, 
\begin{eqnarray}
   \psi ({\vec x}_f, t_f) &=& \int d^3 {\vec x}_i
    \ U ({\vec x}_f, t_f; {\vec x}_i, t_i)
 \ \psi ({\vec x}_i, t_i),
\label{eq:evo1} \\
   \phi ({\vec r}_f, t_f) &=& \int d^3 {\vec r}_i
    \ V ({\vec r}_f, t_f; {\vec r}_i, t_i)
 \ \phi ({\vec r}_i, t_i),
\label{eq:evo2}
\end{eqnarray}
are related by
\begin{eqnarray}
  U ({\vec x}_f, t_f; {\vec x}_i, t_i)
&=& ( a_f \ a_i )^{\frac{3}{2}}
  \exp \left[ -i \frac{m}{2} \left( \dot a_f \ a_f \ {\vec x}_f^{\ 2}
   - \dot a_i \ a_i \ {\vec x}_i^{\ 2} \right) \right]
\ V ( a_f {\vec x}_f, t_f; \ a_i {\vec x}_i, t_i),
\label{eq:dual3}\\
  V ( {\vec r}_f, t_f; {\vec r}_i, t_i )
&=& ( a_f \ a_i )^{- \frac{3}{2}}
\exp \left[ \ i \frac{m}{2} \left( h_f \ {\vec r}_f^{\ 2}
                                 - h_i \ {\vec r}_i^{\ 2} \right) \right]
 \ U \left( \frac{{\vec r}_f}{a_f}, t_f;
            \frac{{\vec r}_i}{a_i}, t_i \right).
\label{eq:dual4}
\end{eqnarray}
  Here we adopt a shorthand notation to suppress the time dependence,
$a_i \equiv a(t_i)$, and $\dot a_i \equiv {d a}/{d t}(t_i)$, etc..
  In the following, we shall refer to these transformations,
Eqs.(\ref{eq:dual1}), (\ref{eq:dual2}), (\ref{eq:dual3}), (\ref{eq:dual4}), as
dual transformations.
  Finally, the expectation values of physical observables in either class
of systems follow a similar rule which leaves their values
invariant under dual transformations.
\begin{equation}
\int d^3 x \ \psi^* (\vec x,t) \, {\cal O} \, \psi(\vec x,t) =
\int d^3 r \ \phi^* (\vec r,t) \, {\cal O} \, \phi(\vec r,t).
\end{equation}

\section{Evolution Operator of a Simple Harmonic Oscillator}

  The exact dual transformations, Eqs.(\ref{eq:dual1}), (\ref{eq:dual2}), 
(\ref{eq:dual3}), (\ref{eq:dual4}), can be used to map known solutions in 
one class to another.
  As a special case, the quantum dynamics of a free particle
($V(\vec x,t) = 0$) moving in the RW background metric is equivalent to a 
quantum mechanical simple harmonic oscillator (SHO), Eq.(\ref{eq:sch4}), 
with a time-dependent frequency, 
\begin{equation}
 \omega^2 (t) \equiv - \frac{{\ddot a}(t)}{a (t)}.
\label{eq:freq}
\end{equation}

  To show the use of dual transformation, Eq.(\ref{eq:dual4}), we begin 
with the solution of a free particle in the RW background metric.
  For instance, Eq.(\ref{eq:sch1}) can be solved by redefining a new time
variable,
\begin{equation}
    \frac{d \tau}{d t} = \frac{1}{a^2 (t)}
    \hspace{0.75cm} \Rightarrow \hspace{0.75cm}
    \tau(t) - \tau(t_0) = \int_{t_0}^{t} \frac{d s}{a^2 (s)}.
\label{eq:tau}
\end{equation}
  In this new time variable, the Schr\"{o}dinger equation, Eq.(\ref{eq:sch1}),
becomes 
\begin{equation}
 i \frac{\partial}{\partial \tau} \psi( \vec x, t(\tau) ) =
 - \frac{ {\nabla_x}^2 }{2 m} \psi( \vec x, t(\tau) ),
\end{equation}
and the plane-wave solution is given by
\begin{equation}
    \psi( \vec x, t; \vec k ) = \exp
    \left\{  i \left[\vec k \vec x - \ \omega (\vec k) \tau(t) \right]
    \right\},
 \hspace{1cm} \omega ( \vec k ) \equiv \frac{ {\vec k}^2 }{2 m}.
\label{eq:planewave}
\end{equation}
  From this plane wave solution, one can extract the evolution operator,
Eq.(\ref{eq:evo1}), through the Fourier transform.
  In the coordinate representation, it is given by
\begin{equation}
         U ( {\vec x}_f , t_f ; {\vec x}_i , t_i ) =
    \left\{ \frac{m}{2 \pi i [ \tau(t_f) - \tau(t_i) ]} \right\}^{3/2}
      \exp \left\{ \frac{i \ m \ ({\vec x}_f - {\vec x}_i)^2}
                        {2 \ [ \tau(t_f) - \tau(t_i) ]} \right\}.
\label{eq:free}
\end{equation}

  Given a scale factor $a(t)$ and the evolution operator of a free system,
Eq.(\ref{eq:free}), we can derive the exact evolution operator of the
corresponding system in Class B,
\begin{equation}
 \hspace{-0.4cm} V ( {\vec r}_f, t_f; {\vec r}_i, t_i ) =
\left[ \frac{m C(t_f,t_i)}{2 \pi i} \right]^{3/2}
\exp \left\{ \frac{i m}{2}
     \left[ A(t_f,t_i) {{\vec r}_f}^{\ 2} +
            B(t_f,t_i) {{\vec r}_i}^{\ 2} -
          2 C(t_f,t_i)  {\vec r}_f \cdot {\vec r}_i \right] \right\},
\label{eq:dual5}
\end{equation}
where
\begin{eqnarray}
 & & A(t_f,t_i) \equiv   h(t_f)
          + \frac{1}{a^2 (t_f) [ \tau (t_f) - \tau (t_i) ] },
     \hspace{0.3cm}
     B(t_f,t_i) \equiv - h(t_i)
          + \frac{1}{a^2 (t_i) [ \tau (t_f) - \tau (t_i) ] },
\nonumber \\
\vspace{-1cm}
& & C(t_f,t_i) \equiv
            \frac{1}{a(t_f) a(t_i) [ \tau (t_f) - \tau (t_i) ]}.
\label{eq:dual6}
\end{eqnarray}
  We emphasize that these two evolution operators, Eqs.(\ref{eq:free}), 
(\ref{eq:dual5}) and (\ref{eq:dual6}), give complete solutions to the 
quantum dynamics of both systems; free particle in the RW background metric 
and simple harmonic oscillator with time-dependent frequency, 
Eq.(\ref{eq:freq}).
  In the following, we shall examine several special examples of this
general formula.
  For simplicity, we shall focus on time-independent
$(\frac{d}{d t} \omega^2 (t) = 0)$ cases:

  {\bf Case (1) :  \ $\omega_1^2 (t) = 0
                      \hspace{0.5cm} \Rightarrow \hspace{0.5cm}
                             a_1 (t) = \alpha t + \beta, \hspace{0.5cm}
                             h_1 (t) = {\alpha}/({\alpha t + \beta}).$ }

  In this case, the dual transformation maps a free particle moving in
a linearly expanding space to a free particle in flat space.
  The $\tau$ function, Eq.(\ref{eq:tau}), is given by
\begin{equation}
    \tau_1 (t_f) - \tau_1 (t_i) \equiv
    \int_{t_i}^{t_f} \frac{ds}{a_1^2 (s)}
  = \frac{1}{\alpha ( \alpha t_i + \beta )}
  - \frac{1}{\alpha ( \alpha t_f + \beta )}
  = \frac{T}{ a(t_f) a(t_i) },
\end{equation}
where $T \equiv t_f - t_i$ is the time difference.
  Substituting  these factors into the general solutions for evolution
operators, Eqs.(\ref{eq:dual5}), (\ref{eq:dual6}), we get
\begin{equation}
  A_1 (t_f,t_i) = B_1 (t_f,t_i) = C_1 (t_f,t_i) = \frac{1}{T},
\end{equation}
and the evolution operator is
\begin{equation}
   V_1 ( {\vec r}_f, t_f; {\vec r}_i, t_i ) =
   \left( \frac{m}{2 \pi i T } \right)^{3/2}
   \exp \left[ \frac{i \ m \ ({\vec r}_f - {\vec r}_i)^2}{2 T} \right],
\end{equation}
a free particle propagator as one would expect from Eq.(\ref{eq:sch4}).

  {\bf Case (2) : \ $\omega_2^2 (t) = \omega^2
                \hspace{0.5cm} \Rightarrow \hspace{0.5cm}
               a_2 (t) = A \cos ( \omega t - \delta ), \hspace{0.5cm}
               h_2 (t) = - \omega \tan ( \omega t - \delta ).$ }

  In this case, the dual transformation maps a free particle moving in
a space with oscillating RW metric to a flat space simple harmonic oscillator.
  The $\tau$ function, Eq.(\ref{eq:tau}), is given by
\begin{eqnarray}
 \tau_2 (t_f) - \tau_2 (t_i) &\equiv&
\int_{t_i}^{t_f} \frac{ds}{a_2^2 (s)}
 = \frac{1}{\omega A^2} [ \ \tan ( \omega t_f - \delta )
                          - \tan ( \omega t_i - \delta ) \ ]
\nonumber \\
  &=& \frac{1}{\omega^2 A^2} [h_2(t_i) - h_2(t_f)].
\end{eqnarray}
  Plugging these formulae into Eq.(\ref{eq:dual6}), and making some 
algebraic simplification, we get 
\begin{equation}
  A_2(t_f,t_i) = B_2(t_f,t_i) = \frac{\omega}{\tan \omega T},
  \hspace{1cm}
  C_2(t_f,t_i) = \frac{\omega}{\sin \omega T}.
\end{equation}
  Thus, the duality solution, Eq.(\ref{eq:dual5}), reduces to the evolution 
operator of a simple harmonic oscillator, 
\begin{equation}
  V_2 ({\vec r}_f, t_f; {\vec r}_i, t_i)
  = \left[ \frac{m \omega}{2 \pi i \sin \omega T}
    \right]^{3/2} \exp
       \left\{ \left( \frac{i m \omega}{2 \sin \omega T} \right)
       \left[ (\cos \omega T) ({{\vec r}_f}^{\ 2} + {{\vec r}_i}^{\ 2})
       - 2 {\vec r}_f \cdot {\vec r}_i \right] \right\}.
\label{eq:sho}
\end{equation}
  We emphasize that this derivation, through the use of an equivalent relation,
is purely algebraic and constitutes the simplest derivation of a SHO 
evolution operator to the best of our knowledge.

 {\bf Case (3) : \ $\omega_3^2 (t) = - \omega^2
              \hspace{0.4cm} \Rightarrow \hspace{0.4cm}
              a_3 (t) = A \cosh (\omega t - \delta), \hspace{0.5cm}
              h_3 (t) = \omega \tanh (\omega t - \delta).$ }

  In this case, the dual transformation maps a free particle moving in an 
inflating (or deflating) space to a simple harmonic oscillator with an
imaginary frequency.
  The $\tau$ function and the evolution operator can be obtained through
analytic continuations of the corresponding results in case (2), and
we shall not give details here.

  We add a side comment considering the nature of this equivalent relation.
  It should be clear from the above discussion that our duality mapping,
in a strict sense, is a homomorphism from Class A models to the Class B ones.
  Many different Class A models get transformed into the same model in
Class B, and the inverse transformation from Class B to Class A is not
well-defined unless we specify the scale factor $a(t)$.
  The scale factors $a(t)$ associated with ``equivalent" systems in Class A
which have identical Class B representation, are specified, through the 
defining relation, Eq.(\ref{eq:freq}), by a set of ``moduli parameters".
  In the ``free-particle vs. SHO" correspondence, these moduli parameters
are linear expansion rate $\alpha$ and initial scale $\beta$ in the first
case, amplitude $A$ and phase shift $\delta$ in the second and third cases.
  Our explicit algebraic calculations of evolution operators using
Eqs.(\ref{eq:dual5}), (\ref{eq:dual6}), verify that the dependence of
these moduli parameters indeed cancels out for the equivalent Class A models.
  This cancellation has its physical origin in the Class A dynamics, as both 
linear expansion rate $\alpha$ and amplitude $A$ amount to a choice of length 
unit, and the change of either initial scale $\beta$ or phase shift $\delta$ 
imply a shift in time.
  Therefore, equivalent Class A systems, in the sense of dual transformation, 
are also physically indistinguishable and the dual transformation gives a 
one-to-one correspondence between physical systems in both Classes.

\section{Moving Soliton Solution from Dual Transformation}

  The dual transformations not only provide us with the connections between 
dynamics of two classes of models, they are also useful in generating 
nontrivial solutions.
  For instance, in the special case of free particle with the scale factor
$a(t) = A \cos (\omega t - \delta)$, we can map the SHO ground state wave 
function, $\phi_0$, from Class B to a localized state in Class A,
\begin{eqnarray}
  \psi_0 ( \vec x, t )
  &=& a^{3/2} (t) \exp \left[ - i \frac{m}{2}
                {\dot a} (t) a (t) {\vec x}^{\ 2} \right] \
  \phi_0 ( a(t) \vec x, t ) \nonumber \\
  &=& \left( \frac{m \omega a^2}{\pi} \right)^{3/4}
   \exp \left[ - i \left( \frac{\omega}{2} \right) t
               - \left( \frac{m}{2} \omega a^2 +
                 i \frac{m}{2} {\dot a} a \right) {\vec x}^{\ 2} \right].
\label{eq:soliton}
\end{eqnarray}

  Since the energy eigenstate in Class B is stationary and in-dissipative,
we expect that the corresponding state in Class A shares the same properties.
  Unlike the wave-packet solution to the non-relativistic free 
Schr\"{o}dinger equation, our localized solution does not disperse in time.
  Indeed, the solitonic solution we found has a finite width in the physical 
length unit, 
\begin{equation}
  \sqrt{\langle [ a(t) \vec x ]^2 \rangle} = {1 \over \sqrt{2 m \omega}}.
\end{equation}
  Finally, the maximum value of the probability density $|\psi|^2$ stays at the
origin $\vec x = 0$, and we conclude that the localized state is a static 
soliton.

  Similarly, there is a corresponding stationary solitonic state in Class A for
each excited state in Class B.
  The properties of these excited solitons modify accordingly.
  For example, when considering a $n$th SHO eigenstate in Class B, the width
of the corresponding soliton increases by a factor $\sqrt {2 n + 1}$.
  It is worth noting that the existence of these soliton solutions is not 
obvious by looking directly at the Schr\"{o}dinger equation in Class A.
  With the aid of dual transformation, we discover these nontrivial solutions 
from the dual Schr\"{o}dinger equation, Eq.(\ref{eq:sch4}).

  Since we start with a homogeneous and isotropic background, the origin of the
soliton can be located at any point $\vec x$.
  Furthermore, the dynamics of non-relativistic free particle systems are
invariant under Galilean transformations.
  Consequently, we can generate a moving soliton by ``boosting" the coordinate
system with a ``velocity" $- \vec v$, 
\begin{equation}
  \vec {x'} \equiv \vec x + \tau (t) \vec v.
\label{eq:trans}
\end{equation}
  The wave function of a moving soliton in the prime coordinate,
\begin{equation}
 \psi_v ( \vec {x'}, t) \equiv \exp \left[ i m \vec v \vec {x'}
  - i \left( \frac{m {\vec v}^{\ 2}}{2} \right) \tau (t) \right] \
      \psi_0 ( \vec {x'} - \tau (t) \vec v, t),
\label{eq:soliton1}
\end{equation}
satisfies the Schr\"{o}dinger equation
\begin{equation}
 i \frac{\partial}{\partial t} \psi_v ( \vec {x'}, t) =
 - \frac{{\nabla}^2_{x'}}{2 a^2(t) m} \psi_v ( \vec {x'}, t).
\end{equation}

  As a consistency check for the dual transformations, Eq.(\ref{eq:dual1}), 
one can obtain the same result by first performing a ``boost" in the Class B 
systems, 
\begin{equation}
 \vec {r'} \equiv \vec r + a(t) \tau(t) \vec v.
\end{equation}
and following a similar procedure as described in Sec.II for deriving the 
boosted SHO wave function,
\begin{eqnarray}
 \phi_v ( \vec {r'}, t)
 &\equiv& \exp \left[ i \frac{\eta (t)}{2} m \vec v \cdot
                 (2 \vec {r'} - a(t) \tau(t) \vec v) \right]
 \phi (\vec {r'} - a(t) \tau(t) \vec v, t) \nonumber \\
      &=& \exp \left[ i \frac{\eta (t)}{2} m \vec v
 \cdot    ( \vec {r'} + \vec r) \right]  \phi (\vec r, t),
\label{eq:boost2} \\
 \eta (t) &\equiv& \frac{d}{d t} (a \tau) = \dot a \tau + a \dot \tau,
\end{eqnarray}
which satisfies the corresponding Schr\"{o}dinger equation
\begin{equation}
 i \frac{\partial}{\partial t} \phi_v ( \vec {r'}, t) =
 \left[ - \frac{{\nabla}^2_{r'}}{2 m} - \frac{m}{2}
   \left( \frac{\ddot a}{a} \right) {\vec {r'}}^2 \right] \
     \phi_v (\vec {r'}, t).
\end{equation}
  We can then apply the dual transformation, Eq.(\ref{eq:dual1}), to map the
boosted SHO wave function $\phi_v$ into the ``freely" moving soliton
solution,
\begin{equation}
  \psi_v ( \vec {x'}, t) = a^{3/2} (t)
  \exp \left[ - i \frac{m}{2} \dot a (t) a (t) {\vec {x'}}^2 \right] \
  \phi_v (a (t) \vec {x'}, t).
\label{eq:soliton2}
\end{equation}

  Working out the algebra, one can be convinced that these two procedures,
Eqs.(\ref{eq:soliton1}),(\ref{eq:soliton2}), give the same answer, and we get
\begin{equation}
  \psi_v ( \vec {x'}, t) = \left( \frac{m \omega a^2}{\pi} \right)^{3/4}
   \exp \left[ - i \Phi_v ( \vec {x'}, t )
               - \left( \frac{m}{2} \omega a^2 \right)
                 ( \vec {x'} - \tau \vec v )^2 \right],
\label{eq:moving}
\end{equation}
where the phase factor $\Phi_v ( \vec {x'}, t )$ is given by,
\begin{equation}
 \Phi_v ( \vec {x'}, t ) \equiv \left( \frac{\omega}{2} \right) t
                  + \left( \frac{m {\vec v}^{\ 2}}{2} \right) \tau(t)
                  - m \vec v \cdot \vec {x'}
              + \frac{m}{2} {\dot a} a ( \vec {x'} - \tau \vec v)^2.
\label{eq:phase}
\end{equation}
  In addition, by changing the coordinate $\vec {x'}$ into $\vec x$ in
Eqs.(\ref{eq:moving}), (\ref{eq:phase}), i.e., taking an active view for the
soliton motion, and superimposing solitonic solutions with different 
velocities, we can generate multi-soliton solutions of this system.
  While the linearity property of free Schr\"{o}dinger equation ensures that 
scattering among these solitons is trivial, one might consider adding a 
potential $V(\vec x,t)$ and then study how these solitons interact.

\section{What Do We Learn from this Lesson}

  Due to relatively simple structures, the models constructed above can also 
be solved exactly by other means without referring to the dual transformation.
  At first sight, one might think that these cases are of only pedagogical
interest.
  However, our derivation enjoys several advantage in comparison with other 
standard approaches:

  (1) The use of dual transformation in solving the evolution operators does 
not rely on explicit solutions of eigenstate wave functions or special 
identities among Hermite polynomials, as required in the operator method 
\cite{sak}.
  Throughout our derivation, only elementary functions are involved.

  (2) The boundary conditions of the classical solution,
$\vec q (t_f) = {\vec x}_f,\ \vec q (t_i) = {\vec x}_i$, are built in
directly from the free particle evolution operator, Eq.(\ref{eq:free}).
  The evaluations of functional determinant and the classical action,
which are necessary in the path integral approach \cite{fey}, \cite{sch},
boil down to a single ordinary differential equation, Eq.(\ref{eq:freq}),
in our approach.

  (3) In addition, our treatment allows for more general situations, such as
SHO with a time-dependent frequency, and the study of time-dependent 
phase transition.
  The general solution is summarized in our master formulae,
Eqs.(\ref{eq:dual5}), (\ref{eq:dual6}), and we save the labor without attacking
the problems separately.

  (4) Finally, the dual transformation can be used in both directions.
  Not only this dual relation provides a more efficient way to study 
time-dependent quantum dynamics, but also we can apply this approach to study 
evolution of quantum states in an evolving universe and discover some 
interesting features.
  For instance, the existence of soliton solutions for a free particle,
as shown in the previous section, is not obvious without using the dual 
Schr\"{o}dinger equation.

  Moreover, in the special case of interest, namely, ``free particle in the RW
metric" vs. a ``simple harmonic oscillator" dual relation provides several 
interesting features which is in common with more sophisticated version, 
e.g., ADS/CFT duality \cite{mad}:

  (a) This is the simplest example where the effects of gravitation (or 
geometry) in one class of quantum systems are transformed into a ``real" 
interaction in another class of theories through duality.

  (b) We show how the space-time symmetry, namely, the Galilean invariance,
is realized and transformed according to the duality relation.
  That is, one can check that both Eqs.(\ref{eq:soliton1}), (\ref{eq:boost2}) 
form linear realizations of this symmetry acting on the underlying Hilbert-Fock
space.

  (c) Instead of giving connections among coupling parameters, as typical 
strong-weak dualities do, our duality relation maps a scale factor $a(t)$ into  
a SHO frequency $\omega(t)$.
  The fact that free particles moving in a time-dependent background metric has
its dual realization as a purely bound state problem seems to carry the flavor
of a strong-weak duality.
  Whether this duality has any implication to the real systems, e.g., QCD,
might deserve a close examination.

  (d) One might take the view that the quantum dynamics of a free particle
in the RW metric is, in some sense, a mother theory of various quantum
mechanical models.
  In particular, the quantum mechanical simple harmonic oscillator can be
derived from this mother theory upon specifying a suitable scale factor 
$a(t)$.
  We shall reiterate this point further in the appendix and devise dual
relations for deriving a generating functional from the evolution operators.

\section{Summary and Conclusion}

  In this paper, we give an explicit construction of dual transformations
between quantum mechanical systems in the Robertson-Walker background metric 
and quantum dynamics with an induced quadratic background potential.
  The use of such a dual relation is demonstrated by an algebraic derivation of
the evolution operator for a simple harmonic oscillator, and the presence of 
moving solitonic solution in an oscillating universe.
  Finally, we compare our study with other standard approaches in quantum
mechanics and comment on possible implications of this dual relation.

  We thank Je-An Gu and Ho-Chi Tsai for many interesting conversations
and discussion which initiated the present work.
  We also acknowledge Dr. Pei-Ming Ho, Dr. Yeong-Chuan Kao, and Dr. Miao Li 
for giving useful comments and suggestion.
  C.T. Chan is supported by the NSC grant NSC 89-2811-M002-0072.
  C.K. Chua is supported by the NSC grant NSC 89-2811-M002-0039 and
                                          NSC 89-2811-M002-0086.

\newpage

\appendix

\section{Deriving the Generating Functional of Simple Harmonic Oscillator 
         from Dual Transformations}

  In this appendix, we shall combine tricks in the main text and use them to 
derive the generating functional of a simple harmonic oscillator.
  To simplify notation, our discussion will be restricted to one-dimensional 
case.

  The generating functional of a quantum mechanical system is defined as
\begin{equation}
  W [ f(t); x_f, t_f; x_i, t_i ]
\equiv \int_{q(t_i) = x_i}^{q(t_f) = x_f} {\cal D} q(t)
       \exp \left\{ i \int_{t_i}^{t_f} d t \
  [ \ {\cal L} [ q(t), \dot q (t)] - f(t) q(t) \ ] \right\}.
\end{equation}
  It follows from this definition that all correlation functions can be
generated from the functional derivatives of the generating functional,
assuming $t_i < t_1, t_2, ... , t_n < t_f$,
\begin{equation}
    ( i )^n \frac{\delta^n W [f(t); x_f, t_f; x_i, t_i]}
      { \delta  f(t_1) \ \delta f(t_2) \ ... \ \delta f(t_n) }
       \left|_{f(t)=0} \right.
     = \langle x_f, t_f | {\cal T} [ \hat q (t_1) \ \hat q (t_2) \ ... \
       \hat q (t_n) ] | x_i, t_i \rangle,
\end{equation}
and we can interpret the generating functional as an evolution operator of a 
perturbed system with the modified Hamiltonian given by 
$H' (t) \equiv H + f(t) q$.

  In the following, we shall present two routes for deriving the generating 
functional of a simple harmonic oscillator. 
  The first derivation is based on a gauge transformation between generating 
functional and evolution operator of a simple harmonic oscillator, and
the second derivation examines the corresponding connection for a free 
particle, and then use dual transformation to derive the SHO generating 
functional.

\subsection{Gauge Transformation between Generating Functional and Evolution
            Operator for a Simple Harmonic Oscillator}

  To derive the generating functional of a simple harmonic oscillator
from the exact solution of evolution operator, Eq.(\ref{eq:sho}),
one is tempted to make a coordinate shift, $q \equiv r - s(t)$,
and observes that the simple harmonic potential naturally generates a
coupling to the external source $f_0 (t) \equiv m \omega^2 s(t)$ and a
background energy
$\varepsilon_0 (t) \equiv \frac{1}{2} m \omega^2 s^2(t)$,
\begin{equation}
  V(r) = \frac{1}{2} m \omega^2 r^2 \rightarrow
  \frac{1}{2} m \omega^2
  [ q^2 + 2 s(t) q + s^2(t) ] = V(q) + f_0 (t) q + \varepsilon_0 (t).
\end{equation}
  Motivated by this observation, we can generalize Eqs.(\ref{eq:trans}), 
(\ref{eq:soliton1}), and define a gauge transformed wave function,
\begin{equation}
 \varphi_s (q, t) \equiv \exp \left[ - i m \dot s (t) q
              + i \int_{t_0}^{t} \varepsilon (\tau) d \tau \right]
                  \phi (q + s(t), t), \hspace{0.5cm}
\varepsilon (t) \equiv \frac{m}{2} [\omega^2 s^2 (t) - {\dot s}^2 (t)],
\label{eq:dual9}
\end{equation}
which satisfies the Schr\"{o}dinger equation of a driven oscillator,
\begin{equation}
  i \frac{\partial}{\partial t} \varphi_s (q, t) =
  \left[ - \frac{1}{2 m} \frac{\partial^2}{\partial q^2}
         + \frac{1}{2} m \omega^2 q^2 + f(t) q \right]
  \varphi_s (q, t),
\end{equation}
where the external source function (or driving force) is given by
\begin{equation}
 f(t) \equiv m \omega^2 s(t) + m \ddot s(t).
\label{eq:source}
\end{equation}
  Notice that the time derivatives of the classical path $s(t)$, appearing
in the background energy $\varepsilon(t)$ and the external source $f(t)$,
are induced by the gauge transformation, Eq.(\ref{eq:dual9}).

  Since the generating functional is defined as a function of the external
source $f(t)$, we need to invert the relation, Eq.(\ref{eq:source}), and
solve the classical path $s(t)$ in terms of $f(t)$.
  Thus, with a given driving force $f(t)$, the general solution to the
inhomogeneous ordinary differential equation, Eq.(\ref{eq:source}), is,
\begin{equation}
  s(t) = s_\omega (t) + A \cos (\omega t - \delta),
\label{eq:dsho1}
\end{equation}
where the particular solution, $s_\omega (t)$, is given by a convolution of the
external source with the one-dimensional Green's function $G_\omega (t)$,
\begin{equation}
  s_\omega (t) \equiv \frac{1}{m} \int d \tau
  f(\tau) \ G_\omega (t - \tau),
\label{eq:dsho2}
\end{equation}
\begin{equation}
  G_\omega (t) \equiv \int \frac{d \alpha}{2 \pi}
  \frac{e^{i \alpha t}}
  {\omega^2 - \alpha^2 - i \epsilon}
  = \frac{i}{2 \omega} \left[ \theta(t)  e^{- i \omega t}
                            + \theta(-t) e^{i \omega t}   \right].
\label{eq:dsho3}
\end{equation}
  One can then use this relation to rewrite the gauge transformed wave
function, Eq.(\ref{eq:dual9}), in terms of the external source $f(t)$ and
the original SHO wave function $\phi (r, t)$.

  The evolution operator in the presence of an external source,
\begin{equation}
   \varphi_s (q_f, t_f)  = \int d q_i V_s (q_f, t_f; q_i, t_i)
   \varphi_s (q_i, t_i),
\end{equation}
can now be related to the SHO evolution operator,
via Eqs.(\ref{eq:dual2}),(\ref{eq:dual9}), as
\begin{equation}
 V_s (q_f, t_f; q_i, t_i) = \exp \left[ - i m \dot s_f q_f
                                        + i m \dot s_i q_i
            + i \int_{t_i}^{t_f} \varepsilon (\tau) d \tau  \right]
 V (q_f + s_f, t_f; q_i + s_i, t_i).
\label{eq:dual10}
\end{equation}
  Given the exact SHO evolution operator, $V(r_f, t_f; r_i, t_i)$,
Eq.(\ref{eq:sho}), and the general solution for a driven harmonic
oscillator, Eqs.(\ref{eq:dsho1})$\sim$(\ref{eq:dsho3}), we obtain the
explicit form of generating functional via Eq.(\ref{eq:dual10}),
\begin{eqnarray}
  V_s (q_f, t_f; q_i, t_i) &=&
  \sqrt{ \frac{m \omega}{2 \pi i \sin \omega T} }
  \exp \left\{ \left( \frac{i m \omega}{2 \sin \omega T} \right)
  \left[ (\cos \omega T) [ {(q_f + s_f)}^2 + {(q_i + s_i)}^2 ]
  \right.  \right. \nonumber \\
  & &    \hspace{3cm}
         \left. \left. - 2 (q_f + s_f) (q_i + s_i) \right]
         - i m \dot s_f q_f + i m \dot s_i q_i
     + i \int_{t_i}^{t_f} \varepsilon (\tau) d \tau \right\} \nonumber \\
 \vspace{0.25cm}
 &=& \sqrt{ \frac{m \omega}{2 \pi i \sin \omega T} }
   \exp \left\{ \left( \frac{i m \omega}{2 \sin \omega T} \right)
        \left[ (\cos \omega T) (q_f^2 + q_i^2)
        - 2 q_f q_i \right. \right.
        \nonumber \\
   & &  \hspace{3cm}
        \left. \left. - 2 P_\omega (t_f, t_i) q_f
                      - 2 Q_\omega (t_f, t_i) q_i
                      +   R_\omega (t_f, t_i)
        \right] \right\}
\label{eq:sho1}
\end{eqnarray}
where $T \equiv t_f - t_i$ and various coefficient functions are
\begin{eqnarray}
  P_\omega (t_f, t_i) &\equiv& \hspace{0.4cm}
                       \left( \frac{\sin \omega T}{\omega} \right)
                      {\dot s}_f - (\cos \omega T) s_f + s_i
           = \frac{1}{m \omega}
             \int_{t_i}^{t_f} f (\tau) \sin \omega (\tau - t_i) d \tau, \\
  Q_\omega (t_f, t_i) &\equiv& - \left( \frac{\sin \omega T}{\omega}
                                 \right)
                      {\dot s}_i - (\cos \omega T) s_i + s_f
            = \frac{1}{m \omega}
              \int_{t_i}^{t_f} f (\tau) \sin \omega (t_f - \tau) d \tau,\\
  R_\omega (t_f, t_i) &\equiv& (\cos \omega T) (s_f^2 + s_i^2) - 2 s_f s_i
         + \left( \frac{2 \sin \omega T}{m \omega} \right)
           \int_{t_i}^{t_f} \varepsilon (\tau) d \tau \nonumber \\
     &=& - \frac{2}{m^2 \omega^2}
           \int_{t_i}^{t_f} f (\tau)
           \sin \omega (t_f - \tau)
    \left[ \int^{\tau}_{t_i} f (\sigma)
           \sin \omega (\sigma - t_i) d \sigma \right] d \tau .
\label{eq:sho2}
\end{eqnarray}
This is in agreement with the standard textbook result \cite{fey},
\cite{sch}.

\subsection{SHO Generating Functional from Free Particle Evolution Operator}

  While the previous derivation, making use of a gauge transformation between
theories with and without coupling to an external source, Eq.(\ref{eq:dual10}),
is straightforward and has a direct generalization in field theory, it is
possible to avoid complicated algebra by taking a different route.
  Namely, we can solve the generating functional of a free particle in the
RW background metric first, then apply the same old dual transform, i.e.,
Eq.(\ref{eq:dual4}), to derive the generating functional of a simple harmonic 
oscillator.

  With this strategy in mind, we first derive the generating functional of a
free particle in a static space, i.e., $a(t) = 1$.
  Following a similar idea in deriving Eq.(\ref{eq:dual9}), one can show that
the gauge transformed wave function, as induced by the coordinate
shift, $y \equiv x - z(\tau)$,
\begin{eqnarray}
 \chi_z (y, \tau) \equiv \exp \left[ - i m \dot z (\tau) y
              - i \frac{m}{2} \int_{\tau_0}^{\tau} {\dot z}^2 (\eta)
          d \eta \right]
                  \psi (y + z(\tau), t(\tau)),
\label{eq:dual11}
\end{eqnarray}
satisfies the Schr\"{o}dinger equation with an external source
$j(\tau) \equiv m \ddot z (\tau)$,
\begin{equation}
  i \frac{\partial}{\partial \tau} \chi_z (y, \tau) =
  \left[ - \frac{1}{2 m} \frac{\partial^2}{\partial y^2} + j(\tau) y
  \right] \chi_z (y, \tau).
\label{eq:sch10}
\end{equation}
  The evolution operator in the presence of an external source, or 
equivalently, the generating functional of a free particle,
\begin{equation}
   \chi_z (y_f,\tau_f)  = \int d y_i \
      U_z (y_f,\tau_f; y_i,\tau_i) \chi_z (y_i,\tau_i),
\end{equation}
can be related to the source-free evolution operator, Eq.(\ref{eq:evo1}),
via Eq.(\ref{eq:dual11}),
\begin{eqnarray}
 & & U_0 (y_f, \tau_f; y_i, \tau_i) \equiv
       U (y_f, t(\tau_f); y_i, t(\tau_i)), \\
 & & U_z (y_f, \tau_f; y_i, \tau_i) \nonumber \\
 &=& \exp \left[ - i m \left( \dot z_f y_f - \dot z_i y_i
                  + \frac{1}{2} \int_{\tau_i}^{\tau_f}
                    {\dot z}^2 (\tau) d \tau  \right) \right]
 U_0 (y_f + z_f, \tau_f; y_i + z_i, \tau_i) \nonumber \\
 &=&  \sqrt{ \frac{m}{2 \pi i {\cal T}} }
  \exp \left\{ \frac{i m}{2 {\cal T}} (y_f + z_f - y_i - z_i)^2
                - i m \dot z_f y_f + i m \dot z_i y_i
   - i  \frac{m}{2} \int_{\tau_i}^{\tau_f}
        {\dot z}^2 (\tau) d \tau \right\}
\nonumber \\
 \vspace{0.25cm}
 &=& \sqrt{ \frac{m}{2 \pi i {\cal T}} }
   \exp \left\{ \frac{i m}{2 {\cal T}} \left[ (y_f - y_i)^2
        - 2 P_0 (\tau_f, \tau_i) y_f
        - 2 Q_0 (\tau_f, \tau_i) y_i
        +   R_0 (\tau_f, \tau_i) \right] \right\},
\label{free1}
\end{eqnarray}
where ${\cal T} \equiv \tau_f - \tau_i$ and various coefficients of
$y_f, y_i$ are given by
\begin{eqnarray}
  P_0 (\tau_f, \tau_i) &\equiv& \ \ {\cal T} {\dot z}_f - (z_f - z_i)
  = \frac{1}{m} \int_{\tau_i}^{\tau_f} j (\xi) (\xi - \tau_i) \ d \xi,
  \label{free2}\\
  Q_0 (\tau_f, \tau_i) &\equiv&   - {\cal T} {\dot z}_i - (z_f - z_i)
  = \frac{1}{m} \int_{\tau_i}^{\tau_f} j (\xi) (\tau_f - \xi) \ d \xi,
  \label{free3}\\
  R_0 (\tau_f, \tau_i) &\equiv&  (z_f - z_i)^2 - {\cal T} \int_{\tau_i}^{\tau_f}
           {\dot z}^2 (\xi) d \xi \nonumber \\
     &=& - \frac{2}{m^2}
           \int_{\tau_i}^{\tau_f} j (\xi) (\tau_f - \xi)
    \left[ \int_{\tau_i}^{\xi} j (\eta) (\eta - \tau_i)
       \ d \eta \right] d \xi .
  \label{free4}
\end{eqnarray}

  To derive the generating functional of a free particle in the RW background
metric, we change the time variable from $\tau$ to $t$, via Eq.(\ref{eq:tau}).
  In so doing, the free particle Sch\"{o}dinger equation, Eq.(\ref{eq:sch10}),
becomes 
\begin{equation}
 i \frac{\partial}{\partial t} \chi_z (x, \tau(t)) =
  \left[ - \frac{1}{2 a^2(t) m} \frac{\partial^2}{\partial x^2}
   + \frac{j(\tau(t))}{a^2(t)} x \right]
  \chi_z (x, \tau(t)),
\end{equation}
and the evolution operator $U_z(y_f, \tau_f; y_i, \tau_i)$, or, the 
generating functional in the presence of an external source, 
$\frac{j(\tau(t))}{a^2(t)}$, modifies accordingly.

  To arrange the formula in a canonical form, Eq.(\ref{eq:sch1}), and make 
connection with the previous result, we further redefine the external source,
\begin{equation}
  f (t) \equiv \frac{j(\tau(t))}{a^3(t)},
\end{equation}
and treat the external source coupling as a potential, 
$V(a(t) x, t) = f(t) a(t) x$, in the RW Schr\"{o}dinger equation, 
Eq.(\ref{eq:sch1}).
  Finally, after substituting the generating functional of a free particle in 
the RW metric, Eqs.(\ref{free1})$\sim$(\ref{free4}) with 
$\tau\rightarrow\tau(t)$, into the dual relation, Eq.(\ref{eq:dual4}),
\begin{equation}
 V_s (r_f, t_f; r_i, t_i) = (a_f a_i)^{-1/2}
     \exp \left[ i \frac{m}{2} ( h_f r_f^2 - h_i r_i^2) \right]
 U_z \left( \frac{r_f}{a_f}, t_f; \frac{r_i}{a_i}, t_i \right),
\end{equation}
  we recover the same result for the generating functional of a simple harmonic
oscillator, Eq.(\ref{eq:sho1})$\sim$Eq.(\ref{eq:sho2}).


\references

\bibitem{mad}
J.~Maldacena,
``The large N limit of superconformal field theories and supergravity,''
Adv.\ Theor.\ Math.\ Phys.\  {\bf 2}, 231 (1998),
[hep-th/9711200].

\bibitem{Duff}
M.~J.~Duff, R.~R.~Khuri and J.~X.~Lu,
Phys.\ Rept.\  {\bf 259}, 213 (1995) [hep-th/9412184].

\bibitem{cos}
E.W.~Kolb and M.S.~Turner,
``The Early Universe,''
{\it Redwood City, USA: Addison-Wesley (1990)}.

\bibitem{pqm}
S.~Fl\"{u}gge,
``Practical Quantum Mechanics I,''
{\it Berlin; New York: Springer-Verlag, (1971)}.
Sec.13, p.19.

\bibitem{sak}
J.J.~Sakurai,
``Modern Quantum Mechanics,''
{\it Redwood City, USA: Addison-Wesley (1994)}.
Sec.2.5, p.112.

\bibitem{fey}
R.P.~Feynman and A.R.~Hibbs,
``Quantum Mechanics and Path Integrals,''
{\it  McGraw-Hill. (1965)}.
Sec.3.6, p.64.

\bibitem{sch}
L.S.~Schulman,
``Techniques and Applications of Path Integration,''
{\it  John-Wiley \& Sons. (1981)}.
Sec.6, p.38.
\end{document}